\documentclass[english,prl,aps,reprint,longbibliography,superscriptaddress]{revtex4-1}
\usepackage[T1]{fontenc}
\usepackage[latin9]{inputenc}
\usepackage{amstext}
\usepackage{graphicx}
\usepackage{esint}
\usepackage{babel}
\begin{document}
\title{Metallic Quantized Anomalous Hall Effect without Chiral Edge States}
\author{Kai-Zhi Bai}
\affiliation{Department of Physics, The University of Hong Kong, Pokfulam Road,
Hong Kong, China}
\author{Bo Fu}
\affiliation{School of Sciences, Great Bay University, Dongguan, China}
\author{Zhenyu Zhang}
\affiliation{Hefei National Laboratory of Physical Sciences at the Microscale,
University of Science and Technology of China, Hefei, Anhui 230026,
China}
\author{Shun-Qing Shen}
\email{sshen@hku.hk}

\affiliation{Department of Physics, The University of Hong Kong, Pokfulam Road,
Hong Kong, China}
\date{\today}
\begin{abstract}
The quantum anomalous Hall effect (QAHE) is a topological state of
matter with a quantized Hall resistance. It has been observed in some
two-dimensional insulating materials such as magnetic topological
insulator films and twisted bilayer graphene. These materials are
insulating in the bulk, but possess chiral edge states carrying the
edge current around the systems. Here we discover a metallic QAHE
in a topological insulator film with magnetic sandwich heterostructure,
in which the Hall conductance is quantized to $e^{2}/h$, but the
longitudinal conductance remains finite. This effect is attributed
to the existence of a pair of massless Dirac cones of surface fermions,
with each contributing half of the Hall conductance due to quantum
anomaly. It is not characterized by a Chern number and not associated
to any chiral edge states. Our study offers novel insights into topological
transport phenomena and topological metallic states of matter.
\end{abstract}
\maketitle

\paragraph*{Introduction-}

QAHE is a quantum transport phenomenon in two-dimensional ferromagnetic
materials where the Hall resistance is quantized to the von Klitzing
constant $h/e^{2}$ while the longitudinal resistance disappears \citep{Haldane1988prl,yu2010quantized,qiao2010quantum,chu2011surface,liu2016quantum,ChangCZ23rmp,Shen-book-2nd}.
The materials are band insulators in the bulk, and possess chiral
edge states carrying a dispationless electric current around the system
boundary \citep{halperin1982quantized,Halperin2020rmp}. The electronic
band structures of the materials are characterized by the Chern number \citep{Thouless1982prl,niu1985quantized},
which equals the number of chiral edge states \citep{hatsugai1993chern}.
Over the last decade the effect has been observed experimentally in
a series of topological insulator (TI) films and two-dimensional materials
\citep{chang2013experimental,chang2015high,checkelsky2014trajectory,kou2014scale,Okazaki2022np,li2019intrinsic,deng2020quantum,otrokov2019prediction,liu2020robust,gong2019experimental,chen2019temqahe,chen2019intrinsic}.
The picture of the chiral edge states are also confirmed experimentally
\citep{liu2016large,yasuda2017quantized}. Recently the half-quantized
Hall conductance was reported in a magnetic doped TI film \citep{mogi2022experimental}.
The power-law decay of the Hall current indicates possible existence
of distinct QAHE, which is not characterized by the Chern number or
chiral edge state \citep{zou2022half,zou2023half,fu2022quantum}. This provides
a possible route to explore novel types of QAHE.

\begin{figure}
\includegraphics[width=8cm]{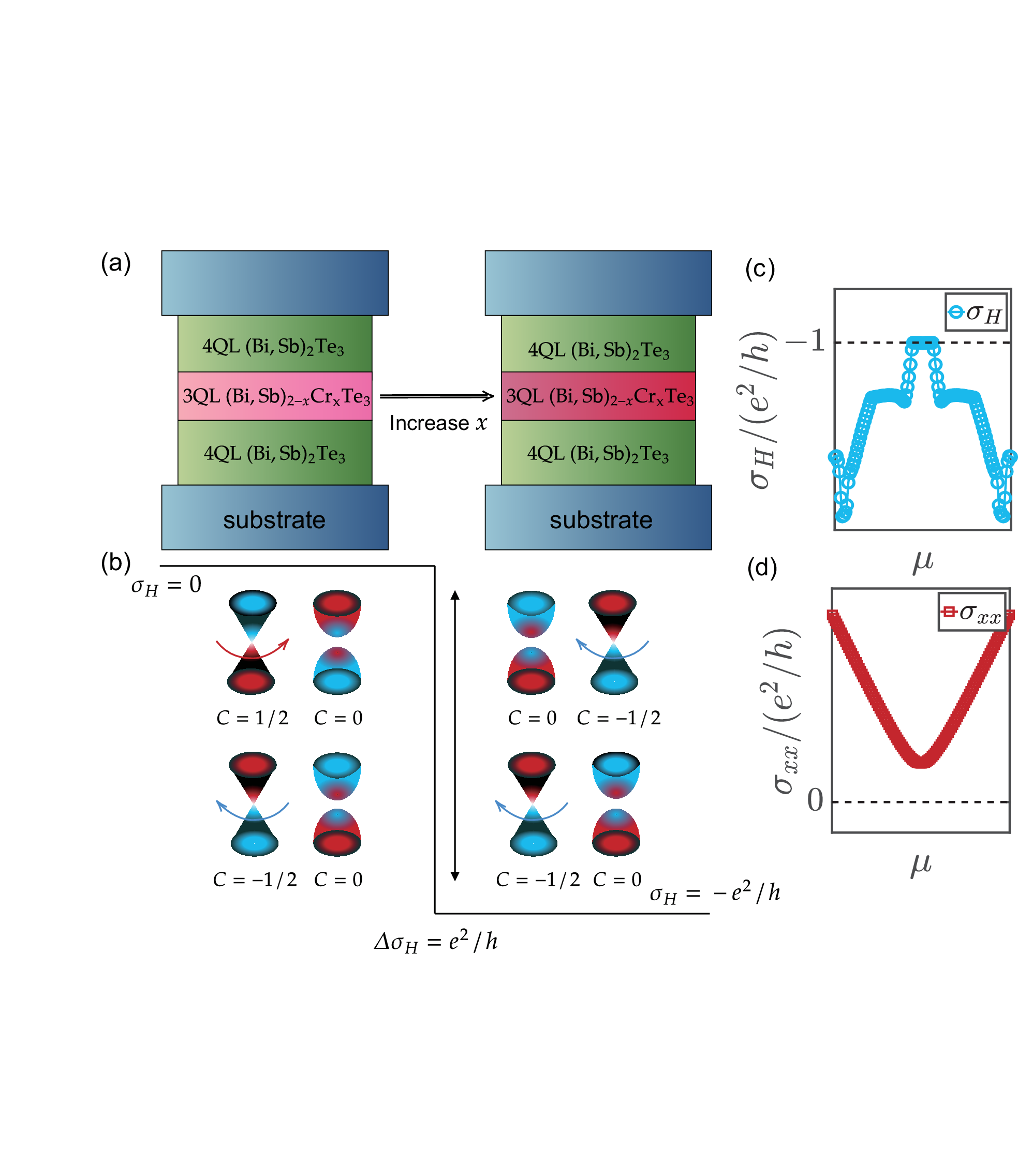}

\caption{(a) Schematic of the magnetic sandwich heterostructure of a $\mathrm{(Bi,Sb)_{2}Te_{3}}$
TI film with the concentration $x$ of magnetically doped Cr atoms.
(b) A transition from two pairs of massless and massive Dirac fermions
with no net Hall conductance $\sigma_{H}=0$ at low concentration
$x$ to that with a quantized Hall conductance $\sigma_{H}=-\frac{e^{2}}{h}$
at higher concentration $x$ (the sign depending on the direction
of magnetization). $C$ represents the Hall conductance in the unit
of $e^{2}/h$, while color represents the sign-value of the Berry
curvature with blue for minus and red for positive. The masses of
a pair of massless and massive Dirac fermions (at the upper horizontal
row) at lower energy exchange by increasing the concentration $x$
while the higher energy parts of the Dirac fermions remain almost
unchanged. (c) Schematic of the quantized Hall conductance $\sigma_{xy}$
and (d) the longitudinal conductivity $\sigma_{xx}$ as function of
the chemical potential $\mu$ at a higher doping concentration $x$.}
\end{figure}

A TI film hosts a pair of massless Dirac cones of electrons near the
two surfaces. The exchange interaction of magnetic ions or the ferromagnetic
magnetization breaks time-reversal symmetry and may manipulate the
nature of the surface states \citep{tokura2019magnetic}. Here we
propose a unique type of QAHE with no chiral edge states in a magnetically
doped TI film in which the Hall conductance is quantized to be $e^{2}/h$
while the longitudinal conductance is finite. The Hall resistivity
is then not quantized. The magnetically doped layers are confined
near the center to form a sandwich structure
as illustrated in Fig 1. Based on numerical calculation and analytical
analysis of the film, it is observed that increasing the concentration
$x$ of doped Cr atoms or increasing the Zeeman field may induce a transition of the
Hall conductance from 0 to $-e^{2}/h$ meanwhile the band structure
shows that no energy gap opens as the magnetically doped layer is
far away from the top and bottom surfaces. Further analysis shows
that the TI film hosts a pair of massless Dirac fermions, one carries
$e^{2}/2h$, and another carries $-e^{2}/2h$ of the Hall conductance
in the absence of the Zeeman field. An increasing Zeeman field drives
one of the gapless Dirac cones and an accompanying gapped Dirac cone
to exchange their masses, and the sign of Hall conductance changes from $e^{2}/2h$ to $-e^{2}/2h$. Consequently,
the total Hall conductance becomes $-e^{2}/h$ (the sign is determined
by the direction of the Zeeman field). The longitudinal conductance
is finite as no gap opens in the surface states, and has a minimal
value when the chemical potential sweeps the Dirac point of the surface
electrons. Hence there do not exist chiral edge states localized near
the system boundary.

\paragraph*{Magnetic sandwich TI film-}

We consider a symmetric TI film with a magnetic doped layer at the
center $m\mathrm{QL}\mathrm{X}_{2}\mathrm{Te}_{3}/3\mathrm{QL}\mathrm{X}{}_{2-x}\mathrm{Cr}_{x}\mathrm{Te}_{3}/m\mathrm{QL}\mathrm{X}{}_{2}\mathrm{Te}_{3}$
with $\mathrm{X}=(\mathrm{Bi},\mathrm{Sb})$ and $m=4$ as shown in
Fig. 1. A larger integer $m$ does not change the main result in this
proposal. $\mathrm{Bi}_{2}\mathrm{Te}_{3}$ and $\mathrm{Sb}_{2}\mathrm{Te}_{3}$
are prototypes of strong TIs  \citep{zhang2009topological}.
$1\mathrm{QL}$ means a quintuple layer of $\mathrm{X}$ and $\mathrm{Te}$
atoms, and is about $1nm$ in $\mathrm{Bi}_{2}\mathrm{Te}_{3}$. The
Dirac cone of surface states was observed explicitly by the ARPES
\citep{chen2009experimental,zhang2010crossover} and was also evidenced
by a series of transport measurements. The exchange interaction between
the p-orbital electron from Bi and Te and magnetic ions Cr may induce
a finite magnetization in $\mathrm{X}{}_{2-x}\mathrm{Cr}_{x}\mathrm{Te}_{3}$\citep{yu2010quantized,tokura2019magnetic}.
Tuning the concentration $x$ of Cr can change the exchange interaction, and even makes it a ferromagnetic insulator
\citep{zhao2020tuning}. The magnetic element Cr was modulation-doped
only near the center layer. The non-doped layers are thick enough
such that the top and bottom surface electrons do not open energy
gap. The topological nature of the band structures of $\mathrm{Bi}_{2}\mathrm{Se}_{3}$
and $\mathrm{Bi}_{2}\mathrm{Te}_{3}$ can be well described by the
tight-binding model for the electrons of $\mathrm{P}_{z,\uparrow}$
and $\mathrm{P}_{z,\downarrow}$ orbitals from $(\mathrm{Bi}$ and
$\mathrm{Te}$ or Se atoms near the Fermi energy \citep{zhang2009topological,liu2010model},

\begin{equation}
H_{TI}=\sum_{l}\Psi_{l}^{\dagger}\mathcal{M}\Psi_{l}+\sum_{l,\alpha=x,y,z}\left(\Psi_{l}^{\dagger}\mathcal{T}_{\alpha}\Psi_{l+\alpha}+\Psi_{l+\alpha}^{\dagger}\mathcal{T}_{\alpha}^{\dagger}\Psi_{l}\right)\label{eq:tight-binding-model}
\end{equation}
where $\mathcal{M}=(m_{0}-2\sum_{\alpha}t_{\alpha})\sigma_{0}\tau_{z}$,
$\mathcal{T}_{\alpha}=t_{\alpha}\sigma_{0}\tau_{z}-i\frac{\lambda_{\alpha}}{2}\sigma_{\alpha}\tau_{x}$,
$\Psi_{l}^{\dagger}$ and $\Psi_{l}$ are the four-component creation
and annihilation operators at position $l=(l_{x},l_{y},l_{z})$. The
Pauli matrices $\sigma_{\alpha}$ and $\tau_{\alpha}$ act on the
spin and orbital indices, respectively. Adapting a model homogeneous
in $x-y$ plane leads to $t_{\parallel}=t_{x}=t_{y}$, $t_{\perp}=t_{z}$,
$\lambda_{\parallel}=\lambda_{x}=\lambda_{y}$, $\lambda_{\perp}=\lambda_{z}$. The magnetic effect
induced by Cr is modeled by introducing the Zeeman
field along the z direction, $V_{Z}=\sum_{l}V_{z}(l_{z})\Psi_{l}^{\dagger}\sigma_{z}\tau_{0}\Psi_{l}$.
$V_{z}(l_{z})=\alpha t_{\perp}$ in the magnetic doped layers (using
$t_{\perp}$ as a unit) with $l_{z}=\pm1/2,\cdots,\pm(m_{z}-1)/2$ where film thickness
$L_{z}$ and the magnetic layer thickness $m_{z}$ are assumed to
be even , and equals zero in the non-doped layers. Here we ignore
the possible change of the bulk gap $m_{0}$ in $\mathrm{X}{}_{2-x}\mathrm{Cr}_{x}\mathrm{Te}_{3}$
caused by doping. 

We consider the periodic boundary condition in the x and y direction. The band structure of the film is calculated numerically by
means of the exact diagonalization method as shown in Fig. 2(a) in
the absence of magnetic layers ($\alpha=0$) and (b) in the presence
of magnetic layers ($\alpha=0.9$). It is observed that there exists
a pair of massless Dirac fermions in both cases. The dispersions are
doubly degenerated near the crossing point at $k=0$. The presence
of the Zeeman field $\alpha$ does not open energy gap in the surface
states while $\alpha$ varies from $0$ to $0.9$. It is reasonable
that the massless surface electrons are mainly located near the top
and bottom surfaces which are far away from the magnetic ions in the
magnetic layers (see Fig. S4 in Ref. \citep{Note-on-SM}). After having
the numerical energy eigenvalues and eigenvectors, the Hall conductance
can be calculated numerically by means of the Kubo formula for electric
conductivity \citep{mahan1981many}. The Hall conductance becomes
nonzero in the presence of $\alpha$ when the Fermi level crosses
the conduction and valence bands with $n>1$. As shown in Fig. 2(c),
a plateau of zero Hall conductance appears near $\mu=0$ for a weak
field, while for a strong Zeeman field $\alpha$, a flat plateau of
$\sigma_{H}=-\frac{e^{2}}{h}$ appears. Detailed calculation presented
in Fig. 2(d) shows the Hall conductance changes from zero to $-\frac{e^{2}}{h}$
with increasing the Zeeman field $\alpha$ for fixed chemical potentials.
Considering that there is no band gap while $\alpha$ changes from
$0$ to $0.9$, the longitudinal conductivity must be finite. Thus the appearance of
the Hall conductance indicates that it differs from the conventional
QAHE in an insulating phase.

\begin{figure}
\includegraphics[width=8cm]{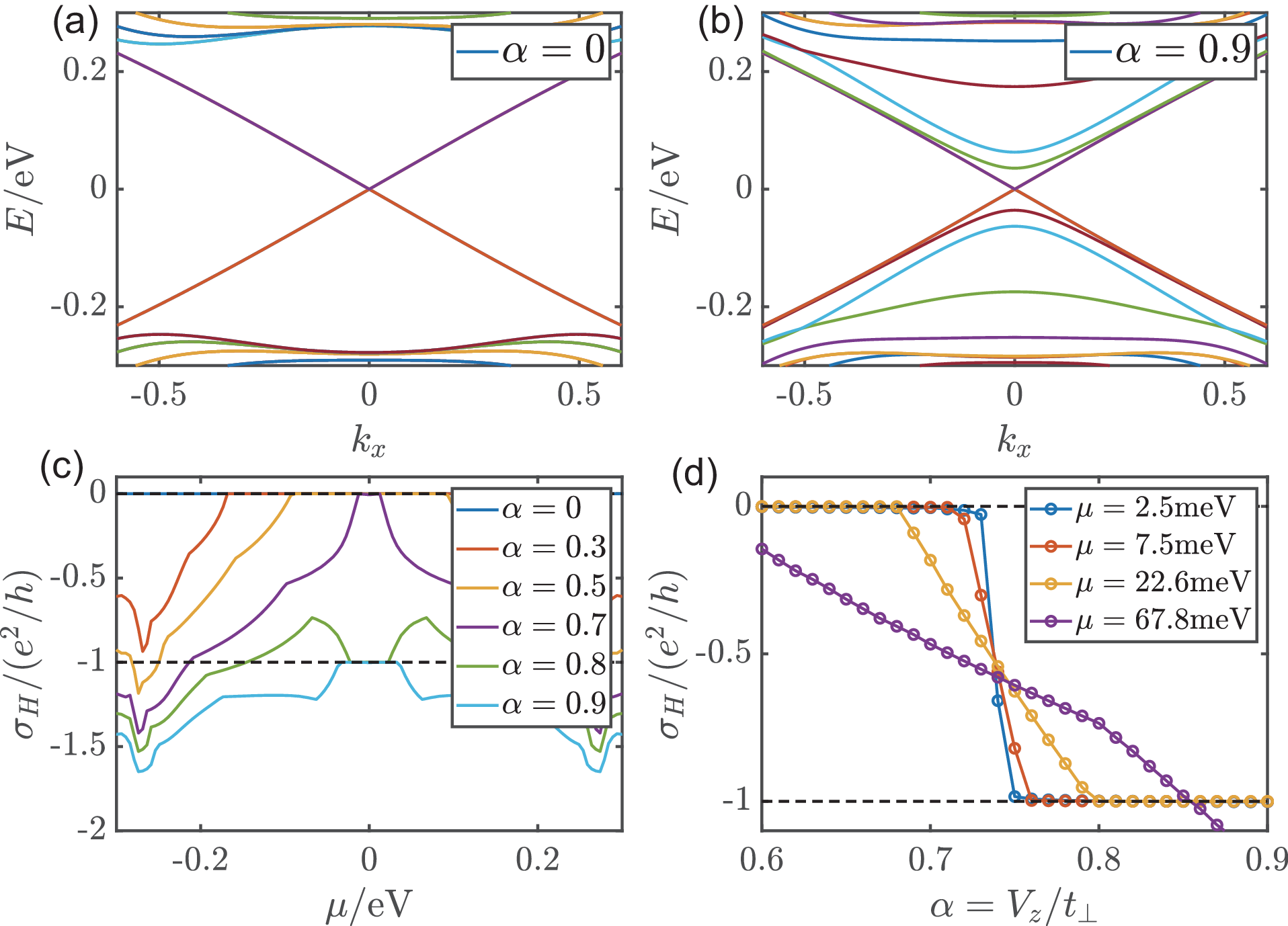}\caption{The band structure near the $\Gamma$ point with $k_{y}=0$ (a) in
the absence of magnetic doping ($\alpha=0$) and (b) in the the presence
of magnetically doping ($\alpha=0.9$). The gapless dispersions for
the surface states in (a) and (b) are doubly degenerated. (c) The
calculated Hall conductance as a function of the chemical potential
$\mu$. (d) The Hall conductance as a function of $\alpha$ at different
chemical potentials. We set the model parameters as $\lambda_{\parallel}=0.41$
eV, $\lambda_{\perp}=0.44$ eV, $t_{\parallel}=0.566$ eV, $t_{\perp}=0.4$
eV, $m_{0}=0.28$ eV, $a=b=1$ nm and $c=0.5$ nm if there is no specific
indication \citep{zhang2009topological}. The thickness $L_{z}=22$
and the magnetic layers $m_{z}=6$. 1QL is about $2c=1nm$.}
\end{figure}

\paragraph*{Equivalent Dirac-like fermions-}

To explore the physical origin of the quantized Hall conductance,
we study the band structure of of the film in the presence of the
Zeeman field. First we adopt the Fourier transformation $\Psi_{l_{z},\mathbf{k}}=\sum_{l_{x},l_{y}}\exp[il_{x}k_{x}+il_{y}k_{y}]\Psi_{l_{x},l_{y},l_{z}}$.
The tight binding model in \ref{eq:tight-binding-model} with the
Zeeman field $H_{tot}=H_{TI}+V_{Z}$ can be split into two parts $H_{tot}=H_{\parallel}+H_{1D}(\alpha)$.
The in-plane spin-orbital coupling $H_{\parallel}=\sum_{l_{z},\mathbf{k}}\Psi_{l_{z},\mathbf{k}}^{\dagger}\lambda_{\parallel}(\sin k_{x}\sigma_{x}+\sin k_{y}\sigma_{y})\tau_{x}\Psi_{l_{z},\mathbf{k}}$.
The part $H_{1D}(\alpha)$ for each $\mathbf{k}$ is equivalent to
a one-dimensional TI with the k-dependent band gap $m(\mathbf{k})=m_{0}-4t_{\parallel}\left(\sin^{2}\frac{k_{x}}{2}+\sin^{2}\frac{k_{y}}{2}\right)$
in a Zeeman field. In the case, $[\sigma_{z},H_{1D}]=0$ such that
$H_{1D}$ can be diagonalized to have a series of energy eigenvalues
$\tilde{m}_{n,\chi}(k_{x},k_{y})$ and eigenvectors $\tilde{\Phi}_{k,n,\chi}=\sum_{l_{z}}U_{n,\chi;l_{z}}\Psi_{l_{z},k}$
with $n=1,...,L_{z}$ and $\chi=\pm$. The double degeneracy is caused
by time-reversal symmetry and inversion symmetry.
Using the eigenvectors as a new basis, we find that $H_{tot}$ is equivalently
reduced to a series of two-dimensional Dirac-like models $H_{tot}\equiv\sum_{\mathbf{k},n,\chi=\pm1}\widetilde{\Phi}_{\mathbf{k},n,\chi}^{\dagger}h_{n,\chi}(\mathbf{k})\tilde{\Phi}_{\mathbf{k},n,\chi}$
with 
\begin{equation}
h_{n,\chi}(\mathbf{k})=\lambda_{\parallel}(\sin k_{x}\sigma_{x}+\sin k_{y}\sigma_{y})+\tilde{m}_{n,\chi}(\mathbf{k},\alpha)\sigma_{z}.\label{eq:Dirac-eq}
\end{equation}
The energy dispersions are $E_{n,\chi,\pm}=\pm\sqrt{\lambda_{\parallel}^{2}(\sin^{2}k_{x}+\sin^{2}k_{y})+\tilde{m}_{n,\chi}^{2}}$
in which $\tilde{m}_{n,\chi}$ plays a role of momentum-dependent
mass term for the Dirac fermions.

In the absence of magnetic doping, i.e., $\alpha=0$, 
$H_{1D}$ can be solved exactly. For details,
the solutions of the energy and wave function can be seen in Ref. \citep{Note-on-SM}. The masses have a relation $\tilde{m}_{n,+}=-\tilde{m}_{n,-}=m_{n}$,
which gives rise to double degeneracy in the band structure rooted in
combination of the time-reversal symmetry and inversion symmetry.
For $m(\mathbf{k})>0$, $H_{1D}$ is topologically nontrivial, and
has zero energy modes $m_{1}=0$; for $m(\mathbf{k})<0$, $H_{1D}$
is topologically trivial, and the lowest energy modes $m_{1}=m(\mathbf{k})$.
Here the film is thick enough such that the finite size effect can
be ignored \citep{lu2010massive}. Therefore, in Eq. (\ref{eq:Dirac-eq}),
$n=1$ corresponds to the pair of gapless bands shown in Fig. 2. The
spatial distribution of the wave function of $m_{1}=0$
is mainly concentrated near the top and bottom surfaces as expected.
The states of nonzero $m_{1}$ or at large $k$ are spatially distributed
in the bulk, which represents that the surface states evolve into
the bulk states with the variation of the wave vector $\mathbf{k}$.
Here the complete band structure
of the gapless Dirac fermions in the entire Brillouin zone consists
of the surface electrons for $m(\mathbf{k})>0$
or small $\mathbf{k}$ and those extended in the z direction for $m(\mathbf{k})<0$
or large $\mathbf{k}$ (see in Fig. 3a). For $n\geq2$, all $m_{n}(\mathbf{k})$
at $\mathbf{k}=0$ are not equal to zero, which means the energy bands
$E_{n,\chi}$ open an energy gap at the point (see Section SI in Ref.
\citep{Note-on-SM}). For a small $\mathbf{k}$, $h_{n,\chi}(\mathbf{k})\simeq\lambda_{\parallel}(k_{x}\sigma_{x}+k_{y}\sigma_{y})+\chi m_{n}(0)\sigma_{z}$.
In other words, all the bands can be regarded as massive Dirac fermions.

In the presence of magnetic doping, the Zeeman field $V_{Z}$ will
change the band structures by altering effective mass $\tilde{m}$,
while linear part vertical to $z$-direction remains unchanged due
to degrees of freedom decoupling. In the basis of the energy eigenstates
of $H_{1D}(\alpha)$ at $\alpha=0$, the Zeeman term can be expressed
as $\alpha\mathbf{I}_{S}(\mathbf{k})\tau_{0}\sigma_{z}$, where $\mathbf{I}_{S}(\mathbf{k})$
is a $L_{z}\times L_{z}$ matrix (see Section SII in Ref. \citep{Note-on-SM})
computable numerically. Thus $H_{1D}$ is projected into the form
$\left(\bigoplus_{n=1}^{L_{z}}m_{n}\tau_{z}+\alpha\mathbf{I}_{S}(\mathbf{k})\tau_{0}\right)\sigma_{z}$,
and further diagonalizing this provides a bijection which maps the
projected Hamiltonian form into the mass term $\oplus_{n}\tilde{m}_{n,\chi}(\mathbf{k},\alpha)\sigma_{z}$.
Confining to the subspace with $\sigma_{z}=+$, we could then track
the evolution and interaction of the mass terms $\tilde{m}_{n,\chi}$
between $n=1$ and $n=2$ blocks with increasing $\alpha$ for given
$\chi$. What stands out in the process is an exotic grafting behavior
signed in Fig. 3: viewing from left to right, while the masses $\tilde{m}_{n,+}(n=1,2)$
maintain their shapes, $\tilde{m}_{n,-}(n=1,2)$, which represent
one massless Dirac cone plus one massive Dirac cone, will fully exchange
their low-energy parts with increasing $\alpha$, i.e., $\textit{massless}\longleftrightarrow\textit{massive}$.
By increasing $\alpha$, $\tilde{m}_{n=1,-}$ and $\tilde{m}_{n=2,-}$
behave as if they cross around $\alpha_{c}\approx0.74$ and then separate,
during which detailed dynamic exchange reveals (see Section SV in
Ref. \citep{Note-on-SM}). On the other hand, what essentially remains
unchanged is the high-energy part of each cone. Then since $\tilde{m}_{n,\chi}$
of $n=1,2$ are naturally assigned with opposite signs for their high-energy
parts, viewing from athelow-energy perspective, their high-energy masses
exchange between massless and massive cones. The induced
mass exchange of the massless and massive Dirac fermions is closely
associated with the sign change of the Hall conductance.

\begin{figure}
\includegraphics[width=8cm]{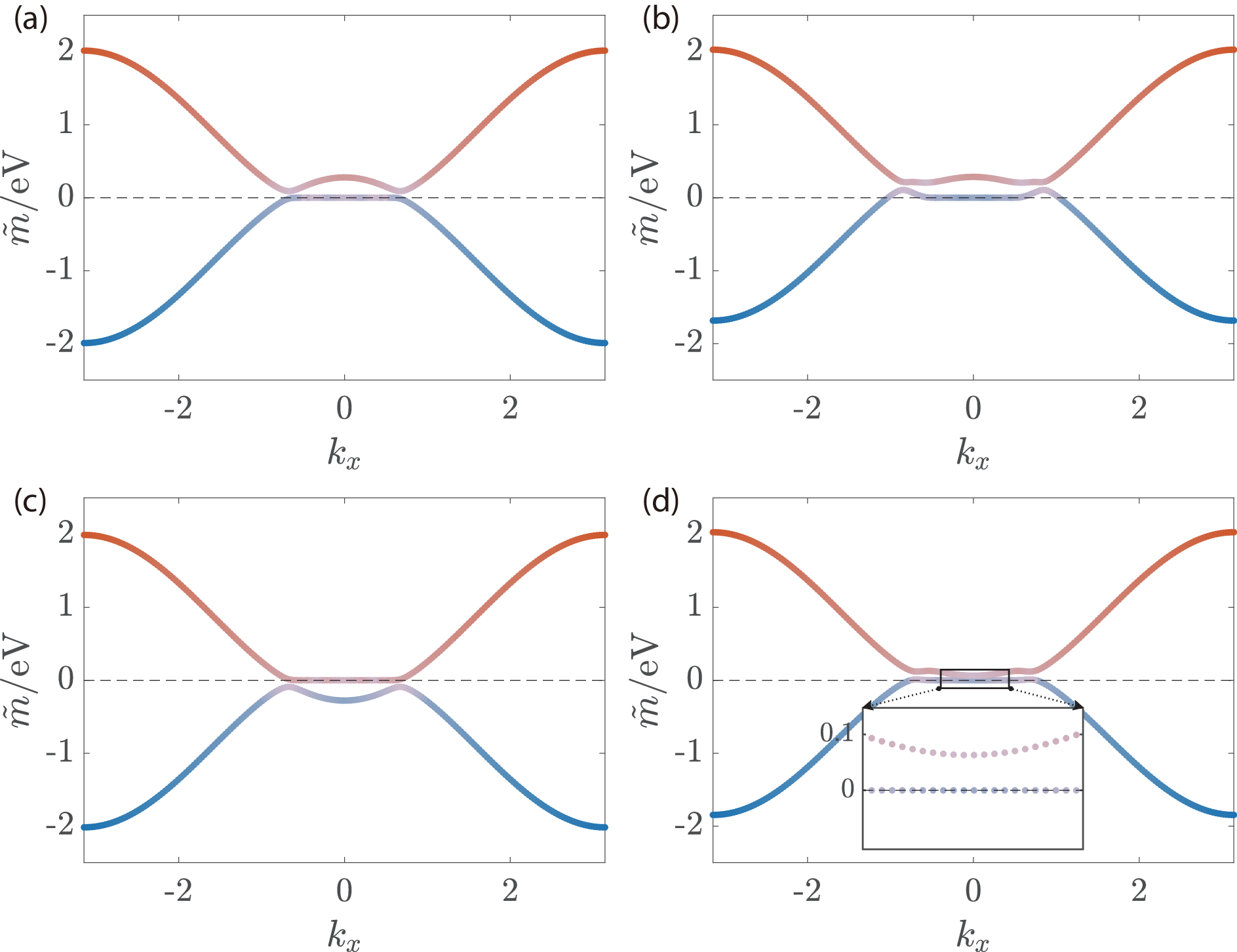}\caption{The evolution of the effective mass $\tilde{m}_{n,\chi}(k_{x},k_{y}=0)$
($n=1,2$). $\tilde{m}_{1,\chi}$ and $\tilde{m}_{2,\chi}$ (a) with
$\chi=+$ at $\alpha$= 0, (b) with $\chi=+$ at $\alpha$= 0.9; (c)
with $\chi=-$ at $\alpha=0$ , (d) with $\chi=-$ at $\alpha=0.9$
.}

\end{figure}

\paragraph*{Quantized Hall conductance-}

The Hamiltonian in Eq. \ref{eq:Dirac-eq}
can be expressed in terms of the spin texture $\mathbf{d}=(\lambda_{\parallel}\sin k_{x},\lambda_{\parallel}\sin k_{y},\tilde{m}_{n,\chi}(k_{x},k_{y}))/E_{n,+}$, $h_{n,\chi}=E_{n,+}\mathbf{d}(\mathbf{k})\cdot\sigma$.
Using the Kubo formula, the Hall conductance
is given by
\begin{equation}
\sigma_{H}=-\frac{e^{2}}{h}\frac{1}{4\pi}\int\frac{dk_{x}dk_{y}}{4\pi}\mathbf{d}\cdot\left[\partial_{k_{x}}\mathbf{d}\times\partial_{k_{y}}\mathbf{d}\right]\left(f_{k,+}-f_{k,-}\right)
\end{equation}
where $f_{k,\pm}=\Theta\left(\mu-E_{n,\pm}\right)$ is the Heaviside
step function for Fermi-Dirac distribution at zero temperature and
$\mu$ is the chemical potential \citep{qi2008topological,Shen-book-2nd}.
For the massive Dirac fermions, the values of $\tilde{m}_{n,\chi}$
at $\mathbf{k}=(0,0)$ and $\mathbf{k}=(\pi,\pi)$ have the same sign,
and there does not exist band inversion in the first Brillouin zone.
The bands are always topologically trivial such that the fully filled
bands, i.e., $\mu=0$, always have no Hall conductance
which is consistent with the TKNN theorem \citep{Thouless1982prl}.
For massless Dirac fermions, $\tilde{m}_{n,\chi}=0$ near $\mathbf{k}=0$.
In the regime, $\mathbf{d}\cdot\left[\partial_{k_{x}}\mathbf{d}\times\partial_{k_{y}}\mathbf{d}\right]=0$
which indicates that the Berry curvature of the band vanishes. Nonzero
Berry curvature comes only from the part of nonzero $\tilde{m}_{n,\chi}$
or the regime of large $k$. The Hall conductance is half-quantized
for $\mu$ located within the regime of $\tilde{m}_{n,\chi}=0$,
$\sigma_{H}=\frac{e^{2}}{2h}sgn[\tilde{m}_{n,\chi}(\pi,\pi)]$. The
quantization is protected by the emergent parity symmetry near the
Fermi surface \citep{fu2022quantum,zou2023half}.

Based on the mass-exchange picture,
we have a theoretical explanation of the change of the Hall conductance
induced by the Zeeman field in Fig. 2(c), (d). The film hosts a series
of massive and massless Dirac fermions. For our purpose, we focus
on the bands of $n=1$ and $n=2$ as all other massive Dirac fermions
($n\geq3$) have no contribution to the Hall conductance when they
are fully filled for the chemical potential near $\mu=0$. In the
absence of the Zeeman field, the film hosts a pair of massless Dirac
fermions, between which one has $+\frac{e^{2}}{2h}$ and the other
has $-\frac{e^{2}}{2h}$ due to the sign difference of the mass terms
at large $k$. The total Hall conductance is zero as expected. The
presence of a weak Zeeman field does not change this situation. Nevertheless,
equipped with a holistic view, when increasing the Zeeman field further,
one massless Dirac fermion and one massive Dirac fermion exchange
their low-energy masses, meanwhile their higher energy parts remain
unchanged, but have different signs. Equivalently, the massless Dirac
fermion changes the sign of massive term at higher energy viewed from
a low-energy perspective. Consequently, its Hall conductance changes
from $+\frac{e^{2}}{2h}$ from $-\frac{e^{2}}{2h}$. During the process, the other massless
Dirac Fermion remains its minus half-quantized Hall conductance unchanged,
and the addition of two massless Dirac fermions gives a quantized
Hall conductance $-\frac{e^{2}}{2h}-\frac{e^{2}}{2h}=-\frac{e^{2}}{h}$.

\paragraph*{Absence of chiral edge states}

There are no chiral edge states around the system boundary in these
paired gapless Dirac fermions. The quantum Hall conductance is not
governed by the Chern number and does not satisfy the conventional
bulk-edge correspondence \citep{hatsugai1993chern}. We calculated
the local density states at the $y$-front surface of a $y$-opened
film in Fig. 4(a), where there is clearly no dispersion that connects
the lateral surface valence and conduction bands, opposite to the
conventional case. This illustrates explicitly that there do not exist
chiral edge states along the system boundary. The asymmetric local
density of states between $k_{x}$ and $-k_{x}$ reflects the fact
that there exists chiral edge current for the filled bulk states.
The states carrying chiral edge current gradually becomes prominent
when immersing into middle of $z$ from its top surface. Furthermore,
it is found that there still exists a chiral edge current whose amplitude
is proportional to the chemical potential due to the time-reversal
symmetry breaking caused by the Zeeman coupling \citep{zou2022half}.
As the Zeeman field is parallel with the lateral surface, the lateral
surface states remain gapless. We present the spatial distribution
of the electric current density in Fig. 4(b). It shows that the current
density is mainly distributed around the surface of the magnetic layers,
and decays quickly into the bulk, which demonstrates that the electronic
transport mainly occurs on the surface. The local current density
on the surface in Fig. 4(c) hows that the current density on the surface
decays slowly which obviously deviates the exponential law. We fit
the numerical result by using the current formula $j_{x}(x)\propto J_{1}(2k_{F}x)/x$
in Ref. \citep{zou2022half}. $J_{1}(x)$
is the first Bessel function. Small deviation appears within expectation
as $\alpha$ is finite, the overall shape which hints a power-law
decay away from ends along $y$ direction, however, is also clear,
as indicated in Fig. 4(c). Also, it is worth of stressing that the
current is induced by the Zeeman exchange interaction, and should
be dispationless. Such behavior depends heavily on the metallic nature
of surface Dirac cone.

\begin{figure}
\includegraphics[width=8cm]{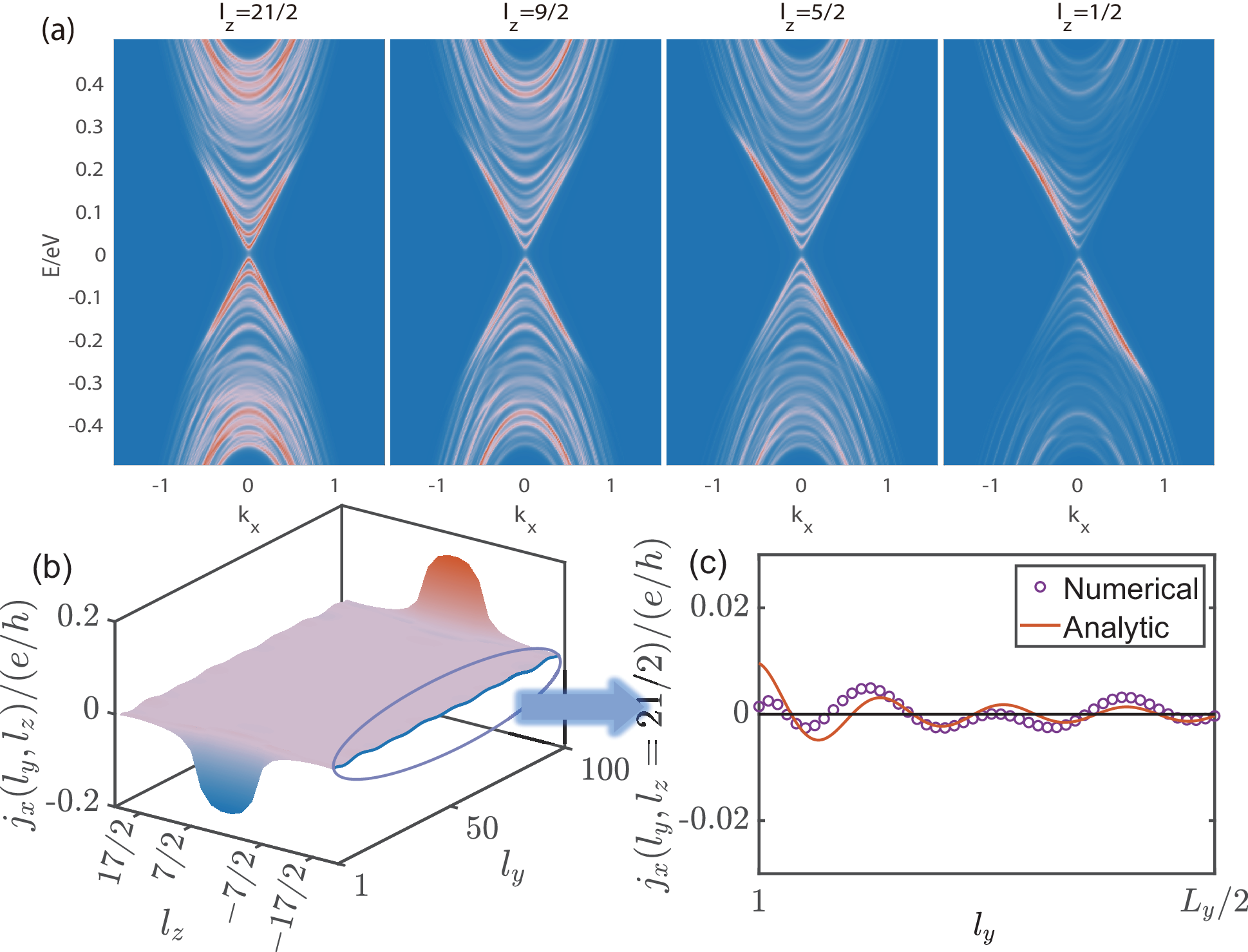}\caption{The local density of states and local current distribution of the
film with $L_{z}=22$, and $\alpha=0.8$. The lattice site $l_{z}=\pm1/2;\cdots;\pm(L_{z}-1)/2$
and $l_{y}=1,\cdots,L_{y}$ (a) Local density of state $\rho(l_{z},k_{x},E)$
at the $y$-front of the film with width $L_{y}=20$. Several representative
$l_{z}$ are picked out to show the absence of chiral edge states.
(b) The local current density $j_{x}(l_{y};l_{z})$ for a film with
width $L_{y}=100$. The chemical potential $\mu=0.1$eV. (c) The surface
current density as a function of $l_{y}$ at the surface $l_{z}=21/2$
or $-21/2$, $j_{x}(l_{y};l_{z}=21/2)$, compared with the current
formula $j_{x}(x)\propto J_{1}(2k_{F}x)/x$ \citep{zou2022half}.}
\end{figure}

\paragraph*{Discussion-}

In the field theory, the massless Dirac fermions possess the parity
symmetry. When the Dirac fermions are coupled to electromagnetic field,
its action fails to restore the symmetry in any regularization, and
is characterized by a half-quantized Hall conductance. The discussion
on parity anomaly in the condensed matter dated back to early 1980s
\citep{Niemi1983prl,Redlich1984prl,Fradkin1986prl}. It has attracted
extensive interests since the discovery of TIs as the massless Dirac
fermions can exist on the surface \citep{zhang2017anomalous,bottcher2019survival,wang2021helical}.
The film here provides a platform to explore the related physics of
parity anomaly. The massless Dirac fermions on the surfaces accompany
with presence of nonzero zero term $\tilde{m}_{n,\chi}$ at large
$k$, which plays a role of the regulators of Dirac fermions in the
field theory. Thus the nonzero Hall conductance is just determined
by the sign of $\tilde{m}_{n,\chi}$ at $k=0$ and large $k$, and
independent of the specific form and the amplitude of $\tilde{m}_{n,\chi}$.
In this sense, the present work reflects the physics of quantum anomaly.
However, we should keep in mind that the term has already broken the
parity symmetry explicitly.
\begin{acknowledgments}
This work was supported by the Research Grants Council, University
Grants Committee, Hong Kong under Grant Nos. C7012-21G and 17301823
and the National Key R\&D Program of China under Grant No. 2019YFA0308603.
\end{acknowledgments}

\end{document}